\documentclass[twocolumn,english,aps,prl,twocolumn]{revtex4-1}

\usepackage[T1]{fontenc}
\usepackage[latin9]{inputenc}
\usepackage{color}
\usepackage{babel}
\usepackage{textcomp}
\usepackage{bm}
\usepackage{amsmath}
\usepackage{amsthm}
\usepackage{amssymb}
\usepackage[unicode=true,pdfusetitle,
 bookmarks=true,bookmarksnumbered=false,bookmarksopen=false,
 breaklinks=true,pdfborder={0 0 0},pdfborderstyle={},backref=false,colorlinks=true]
 {hyperref}
\hypersetup{
 urlcolor=blue, citecolor=blue, linkcolor=blue}

\makeatletter
\theoremstyle{plain}
\newtheorem{thm}{\protect\theoremname}
\theoremstyle{plain}
\newtheorem{prop}{\protect\propositionname}
\theoremstyle{definition}
 \newtheorem{example}{\protect\examplename}

\makeatother

\providecommand{\examplename}{Example}
\providecommand{\propositionname}{Proposition}
\providecommand{\theoremname}{Theorem}

\begin{document}

\title{Genuine activation of nonlocality: From locally available to locally
hidden information }

\author{Somshubhro Bandyopadhyay}
\email{som@jcbose.ac.in, som.s.bandyopadhyay@gmail.com}

\affiliation{Department of Physics, Bose Institute, EN 80, Sector V, Bidhannagar,
Kolkata 700091, India}

\author{Saronath Halder}
\email{saronath.halder@gmail.com}

\affiliation{Harish-Chandra Research Institute, HBNI, Chhatnag Road, Jhunsi, Allahabad
211 019, India}
\begin{abstract}
Quantum nonlocality has different manifestations that, in general,
are revealed by local measurements of the parts of a composite system.
In this paper, we study nonlocality arising from a set of orthogonal
states that cannot be perfectly distinguished by local operations
and classical communication (LOCC). Such a set is deemed nonlocal,
for a joint measurement on the whole system is necessary for perfect
discrimination of the states with certainty. On the other hand, a
set of orthogonal states that can be perfectly distinguished by LOCC
is believed to be devoid of nonlocal properties. Here, we show that
there exist orthogonal sets that are locally distinguishable but without
local redundancy (i.e., they become nonorthogonal on discarding one
or more subsystems) whose nonlocality can be activated by local measurements.
In particular, a state chosen from such a set can be locally converted,
with certainty, into another state, the identity of which can now
only be ascertained by global measurement and no longer by LOCC. In
other words, a locally distinguishable set without local redundancy
may be locally converted into a locally indistinguishable set with
certainty. We also suggest an application, namely, local hiding of
information, that allows us to locally hide locally available information
without losing any part. Once hidden, the information in its entirety
can only be retrieved using entanglement. 
\end{abstract}
\maketitle
\emph{Introduction.} Quantum systems consisting of two or more subsystems
may have nonlocal properties that, in general, are revealed by local
measurements of the parts. Perhaps the most well-known manifestation
of quantum nonlocality, \emph{viz}, Bell nonlocality \citep{Brunner-et-al-2014,Bell-1964,CHSH-1969}
arises from entangled states \citep{Entanglement-horodecki} through
their violation of Bell-type inequalities \citep{Freedman-Causer-1972,Aspect+-1981,Aspect+1982,Hensen+2015,handsteiner+2017,Rosenfeld+,BIG-BELL}.
The latter implies that the predictions of quantum theory cannot be
explained by any local theory. Bell nonlocality is of particular importance
in quantum foundations \citep{Brunner-et-al-2014}, quantum information
\citep{Brunner-et-al-2014} and applications thereof. For example,
Bell nonlocality tests are routinely used to certify device-independent
quantum protocols \citep{Barrett+2005,Acin+2006,Brunner+2008,Pironio+2010,Colbeck-Renner2012}. 

Bell nonlocality, however, is not the only kind of nonlocality of
interest. In this paper, we focus on the nonlocality that arises in
the task of discrimination of quantum states by LOCC \citep{Peres-Wootters-1991,Massar-Popescu-1995,ben99,ben99u,Walgate-2000,Virmani-2001,Ghosh-2001,grois01,walgate-2002,divin03,HSSH,Ghosh-2004,rin04,Watrous-2005,fan-2005,Nathanson-2005,Wootters-2006,Hayashi-etal-2006,nis06,Duan2007,Duan-2009,feng09,Calsamiglia-2010,Bandyo-2011,BGK-2011,Yu-Duan-2012,BN-2013,Cosentino-2013,Cosentino-Russo-2014,B-IQC-2015,Halder+-2019}.
Recall that LOCC protocols are where local observers perform quantum
operations on their respective subsystems and communicate via classical
channels but cannot exchange quantum information. Now suppose that
two or more observers share the parts of a quantum system prepared
in one of several known orthogonal states, the identity of which they
do not know. Their goal is to determine which state the system is
in without error. But because they are separated from each other,
they can only perform measurements realized by LOCC. So the question
here is: Can they perfectly distinguish the orthogonal states by LOCC
as is always possible by a joint measurement on the whole system?
The answer, however, turns out to be no in general. While two orthogonal-pure
states can be perfectly distinguished by LOCC \citep{Walgate-2000},
entangled orthogonal bases, such as the Bell basis, cannot be \citep{Ghosh-2001,HSSH,Ghosh-2004}.
We say that a set of orthogonal states is \emph{locally distinguishable}
if the constituent states can be perfectly distinguished with certainty
by a LOCC protocol; otherwise, \emph{locally indistinguishable}.

A locally indistinguishable set of states is nonlocal in the sense
that a suitable joint measurement on the whole system always yields
more information about the state of the system than any sequence of
LOCC \citep{ben99,grois01,HSSH,Halder+-2019,Chitambar-2013,childs13}.
This new kind of nonlocality has its own share of counterintuitive
results. For example, entanglement is neither necessary nor sufficient
for local indistinguishability. The former is proved by the existence
of orthogonal product states that are locally indistinguishable \citep{ben99,ben99u,nis06,Halder+-2019,Yang13,zhang14,wang15,zhang15,Yang15,xu16-1,Xu-16-2,zhang16,zhang16-1,Xu-17,Wang-2017-Qinfoprocess,Zhang-Oh-2017,zhang17-1,halder}.
They give rise to the so-called ``quantum nonlocality without entanglement''
\citep{ben99} and the recently discovered stronger version of the
same \citep{Halder+-2019}. That entanglement is not sufficient follows
from the result that any two mutually orthogonal pure states are locally
distinguishable \citep{Walgate-2000}. 

Besides exhibiting a different kind of nonlocality, locally indistinguishable
states also imply the existence of locally hidden information: information
encoded in locally indistinguishable states cannot be fully accessed
by the local observers, so part of it remains hidden. For example,
one can encode two classical bits in four Bell states, but only one
bit can be extracted locally simply because the Bell states cannot
be perfectly distinguished by LOCC. The only way to avail the complete
information is by using additional quantum resources such as entanglement.
Quantum cryptography primitives such as data hiding \citep{Terhal2001,DiVincenzo2002,Eggeling2002,MatthewsWehnerWinter}
and secret sharing \citep{Markham-Sanders-2008} rely on this particular
fact. 

For the reasons outlined above, it is not surprising why locally indistinguishable
states have received all the attention so far. That includes finding
new classes \citep{ben99u,Watrous-2005,Duan-2009,BGK-2011,Cosentino-2013,Cosentino-Russo-2014,B-IQC-2015},
extending the formalism to density matrices \citep{Calsamiglia-2010,Bandyo-2011},
obtaining new techniques to prove local indistinguishability \citep{Ghosh-2001,HSSH,Ghosh-2004,fan-2005,Nathanson-2005},
understanding the presence of entanglement \citep{Nathanson-2005,Hayashi-etal-2006}
or lack of it \citep{ben99,ben99u,walgate-2002,childs13}, and more
recently, finding the entanglement cost of distinguishing locally
indistinguishable states \citep{B-IQC-2015}. A set of locally distinguishable
states, on the other hand, is generally understood to be neither interesting
nor important. The reasons being, it is neither nonlocal because the
constituent states can be perfectly distinguished by LOCC nor useful
in ways a locally indistinguishable set can be. In this paper, however,
we will show that this long-held understanding is, at best, incomplete,
for there exist orthogonal states that are locally distinguishable
but deserve consideration on par with their locally indistinguishable
cousins. 

In this paper, we will consider a specific class of LOCC measurements,
namely, orthogonality-preserving-local-measurements (OPLM) \citep{walgate-2002,Halder+-2019}.
The OPLMs are local measurements that keep the post-measurement states
mutually orthogonal, but, on the other hand, they might eliminate
one or more states. Indeed, a LOCC protocol that distinguishes orthogonal
states is a sequence of OPLMs \citep{LOCC-protocol}. We will, of
course, leave out the trivial OPLMs, those that do not change the
states and consider only the nontrivial ones, those where not all
the measurement operators are proportional to the identity. 

Let us now consider the following problem. Suppose Alice and Bob share
a state chosen from a known set $S$ of orthogonal states. They do
not know the identity of the state. They now perform an OPLM $\mathbb{M}$.
Then, for a given outcome $\mu$ of this measurement, they end up
with a state that belongs to a new orthogonal set $S_{\mu}^{\prime}$
whose cardinality $\left|S_{\mu}^{\prime}\right|\leq\left|S\right|$.
Note that the action of an OPLM does not lead to loss of information
\citep{cardinality}. Now, if $S$ is locally indistinguishable, then,
by definition, so is $S_{\mu}^{\prime}$ for all $\mu$. But if $S$
is locally distinguishable instead, can $S_{\mu}^{\prime}$ be locally
indistinguishable? We will make this question more precise in a moment
but before we proceed, let us briefly discuss the issue of local redundancy
\citep{Manik}. 

We say that local redundancy exists in an orthogonal set that remains
orthogonal if we discard one or more subsystems. It may be present
provided at least one of the local dimensions is composite. An example
would make this clear. Let $\left\{ \left|\Psi_{i}\right\rangle \right\} _{i=1}^{4}$
and $\left\{ \left|\Phi_{i}\right\rangle \right\} _{i=1}^{4}$ be
the two-qubit Bell basis and the computational basis respectively.
Consider the set 
\[
\left\{ \left|\Psi_{i}\right\rangle _{\text{AB}}\otimes\left|\Phi_{i}\right\rangle _{\text{A}^{\prime}\text{B}^{\prime}}:i=1,\dots,4\right\} ,
\]
where $\text{A},\text{A}^{\prime}$ and $\text{B},\text{B}^{\prime}$
are the qubit-pairs held by Alice and Bob respectively. First, note
that the above set is locally distinguishable because one can locally
measure $\text{A}^{\prime}\text{B}^{\prime}$ in the computational
basis and correctly learn about the identity of the given state. Now
observe that if we trace out, for example, $\text{AB}$ or $\text{A}^{\prime}\text{B}^{\prime}$
the resulting states remain mutually orthogonal. This is the redundancy
we are talking about, which, however, has consequences. In particular,
discarding $\text{A}^{\prime}\text{B}^{\prime}$ makes the set locally
indistinguishable because Alice and Bob would then share one of the
four Bell states. But, on the other hand, discarding $\text{AB}$
keeps it locally distinguishable. This situation arises only because
of the local redundancy present in the set. For our analysis, therefore,
we will consider orthogonal sets that do not have this redundancy. 

So now we suppose $S$ is locally distinguishable and does not have
local redundancy. As noted earlier, the action of an OPLM $\mathbb{M}$
achieves the following set transformation: $S\rightarrow S_{\mu}^{\prime}$
for the outcome $\mu$, where for every $\mu$ it holds that $\left|S_{\mu}^{\prime}\right|\leq\left|S\right|$.
This brings us to the question that motivated this work: Do there
exist an $S$ and $\mathbb{M}$ such that for any outcome $\mu$,
$S_{\mu}^{\prime}$ is locally indistinguishable? In other words,
does there exist an $S$ such that for any given state chosen from
$S$, Alice and Bob can convert it, with certainty, into a state,
the identity of which can now only be ascertained by a global measurement
and not by LOCC? 

So what we require is the following: For any given input $\rho_{i}\in S$
and any outcome $\mu$ of an OPLM $\mathbb{M}$, $\rho_{i}\rightarrow\sigma_{i}\left(\mu\right)$
such that the orthogonal set $S_{\mu}^{\prime}=\left\{ \sigma_{i}\left(\mu\right)\right\} $
is locally indistinguishable. Note that the requirement cannot be
satisfied if Alice and Bob perform an OPLM that reveals the identity
of the input state (which is possible because the set is locally distinguishable),
or an OPLM whose every outcome results in a definite output. 

An affirmative answer to our question therefore looks improbable and
more so because LOCC operations have inherent limitations \citep{Limitations}.
So it seems safe to conjecture that $S$ should remain locally distinguishable
under OPLMs. However, we will show that this is not the case, in general. 

In particular, we present examples of orthogonal sets from $\mathbb{C}^{2}\otimes\mathbb{C}^{4}$,
$\mathbb{C}^{4}\otimes\mathbb{C}^{4}$, $\mathbb{C}^{5}\otimes\mathbb{C}^{5}$,
and $\mathbb{C}^{5}\otimes\mathbb{C}^{5}\otimes\mathbb{C}^{5}$ with
the following properties: 
\begin{enumerate}
\item Locally distinguishable and without local redundancy. Note that the
local redundancy question does not arise in $\mathbb{C}^{5}\otimes\mathbb{C}^{5}$. 
\item There exists an OPLM that converts the set with certainty into a locally
indistinguishable orthogonal set such that the cardinality remains
unchanged. 
\end{enumerate}
We will also show that not all orthogonal sets have the above two
properties. Therefore, those that do not are genuinely local. Our
result can be viewed as activation of nonlocality by local measurements
in the scenario of quantum state discrimination by LOCC. The activation
is genuine for the sets do not suffer from local redundancy. 

A simple application of our result is the local hiding of locally
available information. This can be understood as follows. We know
the information encoded in a locally distinguishable set is always
locally available. Now suppose it exhibits activable nonlocality.
Then it can be converted into a locally indistinguishable orthogonal
set of the same cardinality by LOCC with certainty. So the information
is no longer completely available locally. This local hiding of information
is irreversible, and to retrieve it in its entirety, one must now
use entanglement. 

Let us first recall a couple of fundamental results in local distinguishability.
We will need them frequently in our proofs.
\begin{thm}
\citep{Walgate-2000} \label{Theorem-1} Two multipartite orthogonal
pure states are locally distinguishable. 
\end{thm}
The next tells us which sets of orthogonal pure states in $\mathbb{C}^{2}\otimes\mathbb{C}^{2}$
are locally distinguishable and which are not. 
\begin{thm}
\citep{walgate-2002} \label{Theorem-2} (a) Three orthogonal pure
states in $\mathbb{C}^{2}\otimes\mathbb{C}^{2}$ are locally distinguishable
iff at least two of those states are product states. (b) Four orthogonal
states in $\mathbb{C}^{2}\otimes\mathbb{C}^{2}$ are locally distinguishable
iff all of them are product states. 
\end{thm}
We begin by considering the simple cases. 
\begin{prop}
\label{prop-1} Two multipartite orthogonal pure states $\left|\varphi_{1}\right\rangle $
and $\left|\varphi_{2}\right\rangle $ remain locally distinguishable
under OPLMs. 
\end{prop}
By Theorem \ref{Theorem-1}, the states are locally distinguishable.
Therefore, an outcome of an OPLM either distinguishes them or converts
them into another orthogonal set which must also contain two pure
states in which case Theorem \ref{Theorem-1} applies. 

Next, consider a locally distinguishable set from $\mathbb{C}^{2}\otimes\mathbb{C}^{2}$. 
\begin{prop}
\label{prop-2} Let $S$ be an orthogonal set of locally distinguishable
states $\left|\varphi_{1}\right\rangle ,\dots,\left|\varphi_{n}\right\rangle $
in $\mathbb{C}^{2}\otimes\mathbb{C}^{2}$, where $2\leq n\leq4$.
Then, under an OPLM, the transformed set remains locally distinguishable. 
\end{prop}
Because of Proposition \ref{prop-1}, it suffices to consider only
the cases: $n=3,4$. Let $n=3$. As the states are locally distinguishable,
at least two of them must be product states {[}Theorem \ref{Theorem-2}(a){]}.
First, suppose that all are product states. Then, for any given outcome
of an OPLM, the new orthogonal states must also be product states
because LOCC cannot convert product states into entangled states.
Then, according to Theorem \ref{Theorem-2}(a), they must be locally
distinguishable. Now suppose two are product states, and the other
is entangled. By a similar argument, the new orthogonal set consists
of either two product states and an entangled state, or three product
states. In both cases, Theorem \ref{Theorem-2}(a) tells us they are
locally distinguishable. Now consider $n=4$. By Theorem \ref{Theorem-2}(b),
all of them must be product states because they are locally distinguishable,
and the previous arguments carry over. 

The above results suggest that, perhaps, locally distinguishable states
do not give up their local distinguishability under OPLMs. Although
compelling, this turns out not to be the case in the higher dimensions. 

\noindent \emph{Orthogonal sets with genuine activable nonlocality.
}Note that, the cardinality of a set of pure states with activable
nonlocality is at least three (follows from Proposition \ref{prop-1}).
We now discuss the examples. 

Let $\left\{ \left|0\right\rangle ,\left|1\right\rangle ,\dots,\left|d-1\right\rangle \right\} $
be an orthonormal basis in $\mathbb{C}^{d}$, where $d\geq2$. Then,
rank-2 and rank-3 projection operators (projectors) are defined as
\begin{eqnarray*}
P_{ij} & = & \left|i\right\rangle \left\langle i\right|+\left|j\right\rangle \left\langle j\right|,i\neq j,\\
P_{ijk} & = & \left|i\right\rangle \left\langle i\right|+\left|j\right\rangle \left\langle j\right|+\left|k\right\rangle \left\langle k\right|,i\neq j\neq k,
\end{eqnarray*}
respectively, where $i,j,k\in\left\{ 0,1,\dots,d-1\right\} $. For
example, $P_{01}=\left|0\right\rangle \left\langle 0\right|+\left|1\right\rangle \left\langle 1\right|$
and $P_{012}=\left|0\right\rangle \left\langle 0\right|+\left|1\right\rangle \left\langle 1\right|+\left|2\right\rangle \left\langle 2\right|$. 
\begin{example}
\label{ex-1} $\mathbb{C}^{2}\otimes\mathbb{C}^{4}$: We assume that
Alice holds a qubit and Bob holds a pair of qubits. We represent the
orthonormal basis corresponding to Bob's state space as follows: $\left|00\right\rangle \equiv\left|\bm{0}\right\rangle $,
$\left|01\right\rangle \equiv\left|\bm{1}\right\rangle $, $\left|10\right\rangle \equiv\left|\bm{2}\right\rangle $,
and $\left|11\right\rangle \equiv\left|\bm{3}\right\rangle $. Consider
the following three orthogonal states (unnormalized):
\begin{eqnarray}
\left|\psi_{1}\right\rangle  & \equiv & \left|0\bm{0}\right\rangle +\left|0\bm{2}\right\rangle +\left|1\bm{1}\right\rangle -\left|1\bm{3}\right\rangle ,\nonumber \\
\left|\psi_{2}\right\rangle  & \equiv & \left|0\bm{0}\right\rangle -\left|0\bm{2}\right\rangle -\left|1\bm{1}\right\rangle -\left|1\bm{3}\right\rangle ,\label{states-ex-1}\\
\left|\psi_{3}\right\rangle  & \equiv & \left|0\bm{1}\right\rangle -\left|1\bm{2}\right\rangle -\left|1\bm{0}\right\rangle -\left|0\bm{3}\right\rangle .\nonumber 
\end{eqnarray}
It is easy to see (and show) that the set does not have local redundancy.
In particular, not all pairs remain orthogonal if we discard any of
Bob's qubits. To show the states (\ref{states-ex-1}) are locally
distinguishable we proceed as follows. First, Alice performs a measurement
on her qubit in the $\left\{ \left|0\right\rangle ,\left|1\right\rangle \right\} $
basis and tells Bob the result. Now, each of Alice's outcome results
in a set of three orthogonal states for Bob to distinguish. If Alice
gets ``$0$'', Bob distinguishes the states $\left|\bm{0}\right\rangle \pm\left|\bm{2}\right\rangle $
and $\left|\bm{1}\right\rangle -\left|\bm{3}\right\rangle $, and
if Alice gets ``$1$'', Bob distinguishes $\left|\bm{1}\right\rangle \mp\left|\bm{3}\right\rangle $
and $\left|\bm{0}\right\rangle +\left|\bm{2}\right\rangle $. 

We now prove the second property. First, Bob performs a binary measurement
defined by the orthogonal projectors $P_{\bm{01}}$and $P_{\bm{23}}$
and informs Alice of the outcome. If Bob gets $P_{\bm{01}}$ they
are left with one of $\left|0\bm{0}\right\rangle \pm\left|1\bm{1}\right\rangle $
and $\left|0\bm{1}\right\rangle -\left|1\bm{0}\right\rangle $. Or,
if Bob gets $P_{\bm{23}}$ they are left with one of $\left|0\bm{2}\right\rangle \mp\left|1\bm{3}\right\rangle $
and $\left|1\bm{2}\right\rangle +\left|0\bm{3}\right\rangle $. So,
in each case, they are left with one of three mutually orthogonal
pure entangled states that can be embedded in a $\mathbb{C}^{2}\otimes\mathbb{C}^{2}$
space. But, according to Theorem \ref{Theorem-2}(a), each set is
locally indistinguishable. So the states given by (\ref{states-ex-1})
can always be locally converted into another set of three orthogonal
states that cannot be locally distinguished. This completes the proof. 
\end{example}
The following example is built on the previous one, \emph{mutatis
mutandis}. But it is interesting in its own right. 
\begin{example}
\label{ex-2} $\mathbb{C}^{4}\otimes\mathbb{C}^{4}$: We assume that
Alice and Bob each holds a pair of qubits. The orthonormal basis corresponding
to each local state space is represented as follows: $\left|00\right\rangle \equiv\left|\bm{0}\right\rangle $,
$\left|01\right\rangle \equiv\left|\bm{1}\right\rangle $, $\left|10\right\rangle \equiv\left|\bm{2}\right\rangle $,
and $\left|11\right\rangle \equiv\left|\bm{3}\right\rangle $. Now,
consider the following orthogonal states (unnormalized):
\begin{eqnarray}
\left|\psi_{1}\right\rangle  & \equiv & \left|\bm{00}\right\rangle +\left|\bm{02}\right\rangle +\left|\bm{31}\right\rangle -\left|\bm{33}\right\rangle ,\nonumber \\
\left|\psi_{2}\right\rangle  & \equiv & \left|\bm{00}\right\rangle -\left|\bm{02}\right\rangle -\left|\bm{31}\right\rangle -\left|\bm{33}\right\rangle ,\nonumber \\
\left|\psi_{3}\right\rangle  & \equiv & \left|\bm{01}\right\rangle -\left|\bm{32}\right\rangle -\left|\bm{30}\right\rangle -\left|\bm{03}\right\rangle .\label{states-ex-2}\\
\left|\psi_{4}\right\rangle  & \equiv & \left|\bm{10}\right\rangle +\left|\bm{12}\right\rangle +\left|\bm{21}\right\rangle -\left|\bm{23}\right\rangle ,\nonumber \\
\left|\psi_{5}\right\rangle  & \equiv & \left|\bm{10}\right\rangle -\left|\bm{12}\right\rangle -\left|\bm{21}\right\rangle -\left|\bm{23}\right\rangle ,\nonumber \\
\left|\psi_{6}\right\rangle  & \equiv & \left|\bm{11}\right\rangle -\left|\bm{22}\right\rangle -\left|\bm{20}\right\rangle -\left|\bm{13}\right\rangle .\nonumber 
\end{eqnarray}
It is a tedious but straightforward exercise to show that the above
set does not have local redundancy. We now show that the states are
locally distinguishable. First, Alice performs a binary measurement
defined by the orthogonal projectors $P_{\bm{03}}$ and $P_{\bm{12}}$
on her qubits. If she gets the first outcome, they end up with one
of the first three states $\left|\psi_{1}\right\rangle ,\left|\psi_{2}\right\rangle ,\left|\psi_{3}\right\rangle $.
Now, these three states are in one to one correspondence with those
given by (\ref{states-ex-1}). This follows by inspection. Hence,
$\left|\psi_{1}\right\rangle ,\left|\psi_{2}\right\rangle \left|\psi_{3}\right\rangle $
are locally distinguishable. Now, if she gets the second outcome,
they end up with one of the three states $\left|\psi_{4}\right\rangle ,\left|\psi_{5}\right\rangle ,\left|\psi_{6}\right\rangle $.
Once again, these are in one-to-one correspondence with those given
by (\ref{states-ex-1}). Hence they are locally distinguishable. So
the whole set is locally distinguishable. 

Now we prove the second property. First, Bob performs a binary measurement
defined by the orthogonal projectors $P_{\bm{01}}$and $P_{\bm{23}}$
and informs Alice of the outcome. If Bob gets $P_{\bm{01}}$ they
are left with one of the six orthogonal states: $\left|\bm{00}\right\rangle \pm\left|\bm{31}\right\rangle $,
$\left|\bm{01}\right\rangle -\left|\bm{30}\right\rangle $, $\left|\bm{10}\right\rangle \pm\left|\bm{21}\right\rangle $,
and $\left|\bm{11}\right\rangle -\left|\bm{20}\right\rangle $. Or,
if Bob gets $P_{\bm{23}}$ they are left with one of another six orthogonal
states: $\left|\bm{02}\right\rangle \mp\left|\bm{33}\right\rangle $,
$\left|\bm{32}\right\rangle +\left|\bm{03}\right\rangle $, $\left|\bm{12}\right\rangle \mp\left|\bm{23}\right\rangle $,
and $\left|\bm{22}\right\rangle +\left|\bm{13}\right\rangle $. Now,
in each case, the corresponding set contains locally indistinguishable
triplets. For example, the first set contains the triplets $\left\{ \left|\bm{00}\right\rangle \pm\left|\bm{31}\right\rangle ,\left|\bm{01}\right\rangle -\left|\bm{30}\right\rangle \right\} $
and $\left\{ \left|\bm{10}\right\rangle \pm\left|\bm{21}\right\rangle ,\left|\bm{11}\right\rangle -\left|\bm{20}\right\rangle \right\} $.
Each triplet is locally indistinguishable by Theorem \ref{Theorem-2}(a)
as the corresponding states can be embedded in a suitable $\mathbb{C}^{2}\otimes\mathbb{C}^{2}$
space. Hence, the whole set must be locally indistinguishable. A similar
argument holds for the second case. So the states given by (\ref{states-ex-2})
can always be locally converted into another set of six orthogonal
states that cannot be perfectly distinguished with certainty by LOCC.
This completes the proof.
\end{example}
By now, the basic idea behind the LOCC protocols proving the desired
properties is clear. In the following example, each local dimensions
is prime and, therefore, there cannot be local redundancy. 
\begin{example}
\label{ex-3} $\mathbb{C}^{5}\otimes\mathbb{C}^{5}$: Consider the
set of three orthogonal states (unnormalized):
\begin{eqnarray}
\left|\phi_{1}\right\rangle  & \equiv & \left|00\right\rangle +\left|11\right\rangle +\left|22\right\rangle +\left|33\right\rangle +\left|44\right\rangle ,\nonumber \\
\left|\phi_{2}\right\rangle  & \equiv & \left|00\right\rangle -\left|11\right\rangle -\left|22\right\rangle -\omega\left|33\right\rangle -\omega^{2}\left|44\right\rangle ,\label{psi-states-4}\\
\left|\phi_{3}\right\rangle  & \equiv & \left|01\right\rangle +\left|23\right\rangle ,\nonumber 
\end{eqnarray}
where $\omega$, $\omega^{2}$ are cubic roots of unity. First we
show the states are locally distinguishable. Alice performs a binary
measurement defined by the orthogonal projectors $P_{02}$ and $P_{134}$
on her system. If she gets the first outcome, they are left to distinguish
$\left|00\right\rangle \pm\left|22\right\rangle ,\left|01\right\rangle +\left|23\right\rangle $.
Now, Bob performs a binary measurement defined by $P_{02}$ and $P_{134}$.
If he gets the first outcome they are left to distinguish the orthogonal
pair $\left|00\right\rangle \pm\left|22\right\rangle $ that we know
can be locally distinguished (Theorem (\ref{Theorem-1})). If he gets
the second outcome they are left only with $\left|01\right\rangle +\left|23\right\rangle $,
so the task is completed. Now, if Alice gets the second outcome $P_{134}$
in the first round, they are left to distinguish a pair of orthogonal
states $\left|11\right\rangle +\left|33\right\rangle +\left|44\right\rangle $
and $\left|11\right\rangle +\omega\left|33\right\rangle +\omega^{2}\left|44\right\rangle $
and Theorem (\ref{Theorem-1}) applies. So we have shown that the
states (\ref{psi-states-4}) are locally distinguishable.

Now we prove the second property. Alice performs a binary measurement
defined by the orthogonal projectors $P_{01}$ and $P_{234}$. There
are only two possible outcomes. If she gets $P_{01}$ they are left
with one the three orthogonal states $\left|00\right\rangle \pm\left|11\right\rangle ,\left|01\right\rangle $.
From Theorem (\ref{Theorem-2})(a) it follows that they are locally
indistinguishable. On the other hand, if she gets $P_{234}$ they
are left with one of the three orthogonal states $\left|22\right\rangle +\left|33\right\rangle +\left|44\right\rangle $,
$\left|22\right\rangle +\omega\left|33\right\rangle +\omega^{2}\left|44\right\rangle $,
and $\left|23\right\rangle $. But we know such a set cannot be perfectly
distinguished with certainty by LOCC \citep{HSSH}. This completes
the proof. 
\end{example}
We now give an example from a multipartite system. Recall that local
(in)distinguishability for a $k$-partite system, where $k\geq3$,
is defined where all the $k$ parties are separated from each other.
However, if a given set is locally indistinguishable in some $k^{\prime}$-partite
configuration, where $k^{\prime}\geq2$, it must be locally indistinguishable.
But note that the converse is false. One can find orthogonal states
in a tripartite system ABC that are locally distinguishable across
all the bipartitions A|BC, B|CA, and C|AB but not in A|B|C \citep{ben99u}. 
\begin{example}
\label{ex-4} $\mathbb{C}^{5}\otimes\mathbb{C}^{5}\otimes\mathbb{C}^{5}$:
Consider the orthogonal states: 
\begin{eqnarray}
\left|\phi_{1}\right\rangle  & \equiv & \left|000\right\rangle +\left|111\right\rangle +\left|222\right\rangle +\left|333\right\rangle +\left|444\right\rangle ,\nonumber \\
\left|\phi_{2}\right\rangle  & \equiv & \left|000\right\rangle -\left|111\right\rangle -\left|222\right\rangle -\omega\left|333\right\rangle -\omega^{2}\left|444\right\rangle ,\label{states-ex-4}\\
\left|\phi_{3}\right\rangle  & \equiv & \left|011\right\rangle +\left|233\right\rangle ,\nonumber 
\end{eqnarray}
The above states are a three-party generalization of those in Example
\ref{ex-3}. The proof is similar. 
\end{example}
\emph{Discussions.} The examples that we discussed provide an idea
of constructions in other state spaces. The first and the second example
may be suitably generalized in $\mathbb{C}^{2n}\otimes\mathbb{C}^{4n}$
and $\mathbb{C}^{4n}\otimes\mathbb{C}^{4n}$ respectively for $n\geq2$.
However, proper care should always be taken to ensure there is no
local redundancy involved when the dimensions of the local subsystems
are composite. The third example could help us to find examples in
other spaces where the local dimensions are prime. However, we do
not know if they could be found in state spaces such as $\mathbb{C}^{2}\otimes\mathbb{C}^{3}$
and $\mathbb{C}^{3}\otimes\mathbb{C}^{3}$. We also showed that sets
with genuine activable nonlocality do not exist in $\mathbb{C}^{2}\otimes\mathbb{C}^{2}$.
Whether there are other state spaces also where they do not exist
is an interesting question. 

Is there an upper bound on the size of sets with genuine activable
nonlocality in a given state-space? We do not have any particularly
helpful intuition here. But we believe other methods for obtaining
such sets could help. So how big they could be in a given state-space
remains an open question. Finally, is it possible to have activable
nonlocality without entanglement? In particular, does there exist
an orthogonal product set with genuine activable nonlocality? We suspect
not. In fact, it would be surprising if it does. 

\emph{Conclusions.} In this paper, we showed that quantum nonlocality
can be genuinely activated in the scenario of quantum state discrimination
by LOCC. In particular, we considered orthogonal sets of pure states
that are locally distinguishable and without local redundancy. We
gave several examples where such a set can be locally converted, with
certainty, into another orthogonal set, which is locally indistinguishable.
That is, a state chosen from an activable set can be locally converted,
with certainty, into another state, the identity of which can be determined
by a global measurement but not by LOCC. We also discussed a potential
application, namely, local hiding of the entire locally available
information. The information, once hidden, is no longer locally available
in full and to access it, one must use entanglement. We also discussed
interesting open questions. 

The notion of activation of quantum nonlocality is known to hold in
the context of Bell-nonlocality \citep{Popescu-1995,Gisin-1996,Peres-1996,Hirsch+2013,Camlet-2017}.
Our result shows that activation of nonlocality also appears meaningfully
in local quantum state discrimination. So whether the activation phenomenon
can also be observed in other manifestations of nonlocality (for example,
Ref. \citep{other-nonlocality}) is an intriguing question. 
\begin{acknowledgments}
We would like to thank Manik Banik, IISER Thiruvananthapuram, for
his observations that led us to address the issue of local redundancy. 
\end{acknowledgments}


\begin{thebibliography}{99}
\bibitem{cardinality} The cardinality of an orthogonal set of states
is related to the amount of classical information. It is obvious that
if $\left|S_{\mu}^{\prime}\right|=\left|S\right|$ there is no loss
of information. On the other hand, if $\left|S_{\mu}^{\prime}\right|<\left|S\right|$
then it implies that we have only extracted part of the classical
information encoded in the parent set $S$.
\bibitem{LOCC-protocol} In a particular round, the local measurements
either identify the state correctly at which point the protocol ends
or failing to do so, preserve orthogonality of the post-measurement
states, possibly eliminating one or more from the contention, so that
the protocol can continue.
\bibitem{Limitations} For example, one could never convert a separable
state into an entangled state, or increase entanglement on an average,
or even reliably distinguish orthogonal product states prepared separately.
\bibitem{other-nonlocality} For example, the recent PBR theorem \citep{Pusey+2012}
and other results \citep{Bandyopadhyay+2014} have shown that there
exist product states for which state elimination with certainty becomes
possible only by entangled measurements on the joint system. The nonlocality
here is in the sense that entangled measurements are necessary for
state elimination with certainty even though the underlying states
are product states.
\end{thebibliography}
\end{document}